\documentclass[preprint,12pt]{elsarticle}

\usepackage{amssymb}
\usepackage{multirow}
\usepackage{multicol}
\usepackage{amsmath}
\journal{Nuclear Physics B}

\begin{document}

%\linenumbers 
\begin{frontmatter}


\title{DBF-Net: A Dual-Branch Network with Feature Fusion for Ultrasound Image Segmentation}

\author[a]{Guoping Xu}
\author[a]{Xiaming Wu}
\author[a]{Wentao Liao}
\author[a]{Xinglong Wu}
\author[a]{Qing Huang}
\affiliation[a]{organization={Hubei Key Laboratory of Intelligent Robotics, School of Computer Science and Engineering, Wuhan Institute of Technology},
            city={Wuhan},
            postcode={430205},
            state={Hubei},
            country={China}}  
\author[b]{Chang Li}       
\affiliation[b]{organization={Department of Biomedical Engineering, Hefei University of Technology},
            city={Hefei},
            postcode={230009},
            state={Anhui},
            country={China}}

\begin{abstract}
 \noindent 
{\bf Objective:} Accurately segmenting lesions in ultrasound images remains a challenging task due to the inherent ambiguity in distinguishing boundaries between lesions and adjacent tissues. Although deep learning has made significant progress in accurate ultrasound image segmentation, there is still a lack of attention to the quality of boundary segmentation and its relationship with the body. As a result, the current accuracy of ultrasound image segmentation still has room for improvement in enhancing the precision of lesion delineation. We aim to improve the overall segmentation accuracy and the quality of boundary segmentation by exploring the relationship between the body and the boundary in ultrasound images.\\
{\bf Methods:} We introduce a dual-branch structure based on a deep neural network that allows the model to learn the relationship between the body and boundary parts under supervision, leading to an improvement in overall segmentation accuracy and boundary segmentation quality. Additionally, we propose a novel feature fusion module aimed at facilitating the integration and interaction of body and boundary information. Based on the dual-branch structure and feature fusion strategy, we have designed UBBS-Net for ultrasound image segmentation.\\
{\bf Results:} The proposed approach is evaluated on three challenging public ultrasound image datasets. Our experimental results demonstrate the superiority of the proposed method compared to existing state-of-the-art methods. Specifically, we achieve a Dice Similarity Coefficient of 81.05$\pm$10.44\% for breast cancer segmentation, 76.41$\pm$5.52\% for brachial plexus nerves segmentation, and 87.75$\pm$4.18\% for infantile hemangioma segmentation on BUSI, UNS and UHES datasets, respectively.\\
{\bf Conclusions:} We propose UBBS-Net, a novel network for ultrasound image segmentation that combines body and boundary information with a proposed feature fusion module. Our method outperforms existing approaches on three challenging public datasets, demonstrating its effectiveness for ultrasound image segmentation. Our code will be available on: https://github.com/apple1986/DBF-Net.

\end{abstract}


% \begin{highlights}
% \end{highlights}


\end{frontmatter}

\section{Introduction}
Image segmentation plays a vital role in medical imaging by enabling physicians to identify and analyze diverse anatomical structures such as tumors, organs, and blood vessels\cite{1}\cite{2}.  Ultrasound (US) imaging ranks among the foremost modalities in medical diagnostics.US imaging employs high-frequency sound waves to provide real-time, non-invasive visualization of internal human body structures.In contrast to other imaging modalities such as computed tomography and magnetic resonance imaging, ultrasound (US) imaging is radiation-free, cost-effective, and portable, and widely available\cite{3}\cite{4}\cite{5}. These advantages make US imaging widely adopted in clinical diagnosis and treatment.

\begin{figure}[!htp]
\centerline{\includegraphics[width=\columnwidth]{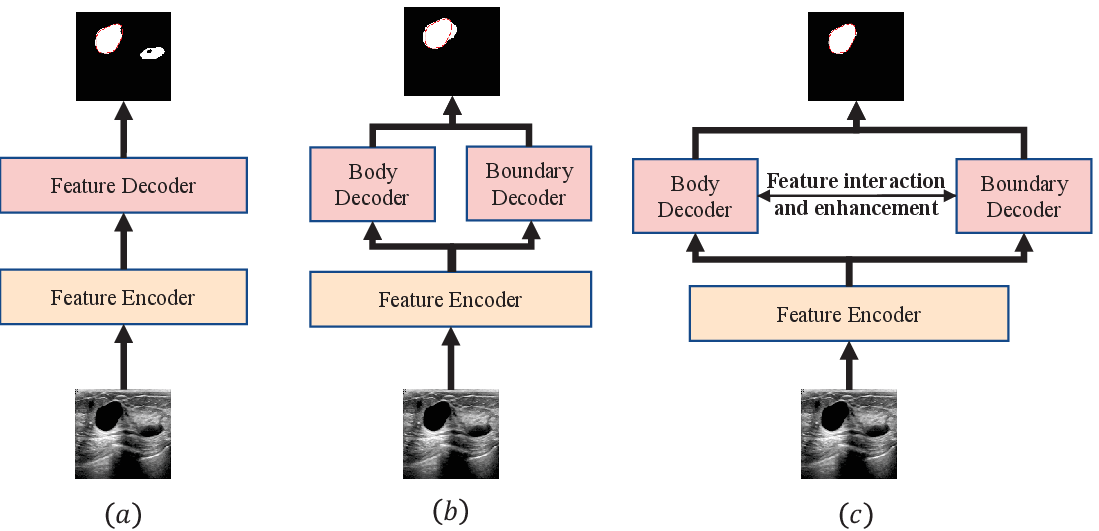}}
\caption{Three different convolution neural network architectures: (a) the classic encoder-decoder structure, (b) a network structure with added boundary information, and (c) the proposed structure in this paper, which do feature interaction and enhancement between body decoder and boundary decoder. The red curve delineates the boundaries of the lesions.}
\label{fig1}
\end{figure}

Semantic segmentation allocates individual pixels within an image to a specific object class. Since the advent of fully convolutional networks (FCN)\cite{6}, the encoder-decoder architecture (Fig. \ref{fig1}(a)) has been widely used for this task, including architectures such as U-Net \cite{7}  and LinkNet \cite{8}. With these advancements, researchers have developed robust and accurate segmentation models that can handle challenges posed by ultrasound images, such as speckle noise, shadows, and artifacts\cite{9}\cite{10}. Recent studies have also shown the efficacy of deep learning methods in segmenting various tissues and organs in ultrasound images, such as breast \cite{10}\cite{11}\cite{12}, thyroid \cite{13}\cite{14}, kidney \cite{15} and liver \cite{17} et al.

Although much progress has been achieved, the segmentation performance still has much space to improve, especially the boundary of the objects or lesions\cite{53}. The shape of the boundary is crucial in clinical diagnosis and treatment, such as judging the benign or malignant lesion.  Despite its importance, boundary segmentation has often been overlooked in prior studies. One of the main reasons is the low resolution of US images, making it difficult to distinguish the boundary between the lesions and nearby tissues or objects. Finally, it is prone to result in over-segmentation or under-segmentation. Therefore, it is critical to enhance boundary segmentation accuracy to achieve optimal segmentation accuracy \cite{18}\cite{19}.

To address the problem mentioned above, some studies have utilized edge information as complementary information to enhance the performance of semantic segmentation (Fig. \ref{fig1}(b)). Typically, Some studies have introduced boundary-aware losses \cite{20}\cite{21} to regulate network training. For instance, a new loss function was devised to achieve balanced weighting between edge and non-edge pixels\cite{20}.Other studies focus on enhancing edge information using additional sub-networks dedicated to edges. For example, HD-Net \cite{22} introduced a two-dimensional boundary decoder that embeds boundary representations to guide the segmentation process.The ET-Net \cite{24}utilizes multiple branches to integrate edge and multi-scale information, thereby improving segmentation accuracy. The RF-Net \cite{11} introduced a technique to improve segmentation accuracy by training residual representations specifically for regions containing challenging-to-predict pixels. DSN-OB \cite{25} employed deep supervision for detecting both boundaries and objects across layers with fine and coarse resolutions. However, those studies ignore the correlation between boundary and body. More precisely, the boundary defines the body part, while the body part delineates the contour of the boundary. Incorporating these relationships into the segmentation model could le ad to a more effective and coherent segmentation pipeline.

Building upon this perspective, we propose a novel approach that integrates ultrasound lesion body segmentation and ultrasound lesion boundary segmentation. Notably, we enable interaction between the body decoder and boundary decoder (Fig. \ref{fig1}(C)). Adhering to the classical encoder-decoder architecture, we design a Feature Fusion and Supervision block (FFS) dedicated to concurrently processing body and boundary information. Specifically, the features of the boundary and body are supervised by the ground truth and fused by the proposed feature fusion block. This approach enables the acquisition of representative feature information, capturing nuances in both the body and boundary aspects. Preliminary findings from this paper were presented at the IEEE International Conference on Bioinformatics and Biomedicine (BIBM) in 2022\cite{38}. However, this paper overlooked the correlation between the boundary and body. To be more precise, it did not incorporate feature interaction and fusion concerning the boundary and body during training. Moreover, the experiments were carried out on a wider range of datasets, leading to significant advancements compared to the previous method. Consequently, the entire approach has undergone substantial transformation.

In summary, the primary contributions of our work are as follows:
\begin{enumerate}

\item We introduce a novel feature interaction module, termed Feature Fusion for Segmentation (FFS), designed to leverage the relationship between body and boundary features for enhanced fusion. 
\item We devise an innovative architecture, named DBF-Net (Dual-Branch Fusion Network), aimed at fusing features from the body and boundary branches for segmenting ultrasound images. We integrate the proposed FFS module to explore the correlation between body and boundary in ultrasound image segmentation.
\item The experimental results, tested in three public US datasets, demonstrate that the proposed method achieves state-of-the-art performance when compared to existing methods.
\end{enumerate}

\begin{figure*}[h]
\centerline{\includegraphics[width=\linewidth]{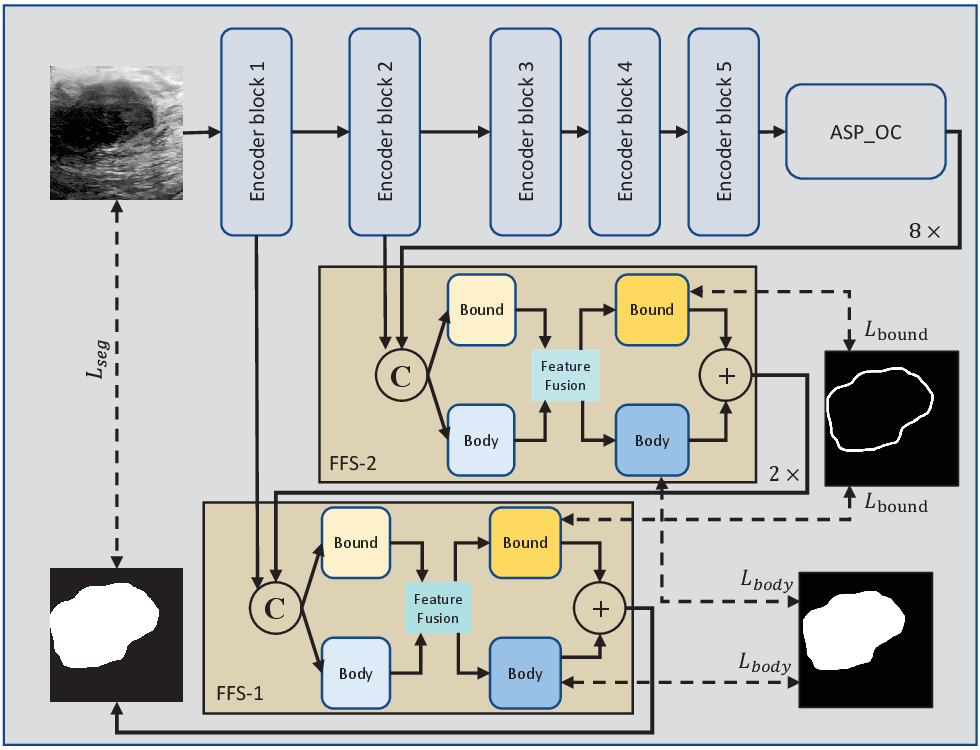}}
\caption{The proposed DBF-Net. 8x and 2x are upsampling ratios.  C and + means concatenation and addition, respectively.  Our model adopts the traditional encoder-decoder architecture.  After the encoder, we utilize  an ASP\_OC module to extract multi-scale features.  The feature fusion and supervision block (FFS) focuses on explicitly modeling body and boundary at the feature level through a set of convolution operations, feature fusion block, and supervision. }
\label{fig2}
\end{figure*}

\section{Methods}
\subsection{Network structure}
As depicted in Fig. \ref{fig2}, our network follows the classic encoder-decoder structure. For the encoder, we use five successive convolution blocks. We denote the output feature representation of each encoder block as $E_{i} \in R^{C\times \left ( H/2^{i-1}  \right ) \times \left ( W/2^{i-1}  \right )} $, where $C$, $H$ and $W$ are channel, input width, and input height, respectively. The $i$ is the $i$-th layer of the model. Each convolution block comprises two convolutional layers, followed by a batch normalization layer and ReLU activation. More detailed information about the encoder can be seen in Table. 1. After the successive convolution block, we use an ASP\_OC  block \cite{34} to capture multi-level information. For the decoder, it consists of two feature fusion and supervision blocks (FFS). ASP\_OC  applies object context pooling and four dilated convolutions, identical to the original ASPP, with rates representing dilation rates. The five output feature maps are then concatenated for the final output. Each convolution is followed by a group of BN and ReLU operations. In FFS, there are two branches that extract body and boundary information from feature maps with deep supervision, respectively. Generally,  The encoder blocks $E_{1}$ and $E_{2}$ focus more on low-level features, like texture and boundary, while $E_{3}\sim E_{5}$  are in favor of semantic information. Therefore, the outputs from encoder blocks $E_{1}$ and $E_{2}$ are inputted into FFS to assist in extracting information about the boundary and body of  segmented objects. 

\begin{table}
\caption{Detailed information of the encoder of DBF-Net}
\centering
\scalebox{0.95}{
\begin{tabular}{c c c c c}
\hline
 Encoder number& kernel size & Dilation & Input channel & Output channel\\ 
 \hline\hline
    1     & 3          & 1     & 3     & 32 \\
    2     & 3          & 2     & 32    & 64 \\
    3     & 3          & 3     & 64    & 128 \\
    4     & 3          & 5     & 128   & 256 \\
    5     & 3         & 7     & 256   & 256 \\
\hline
\end{tabular}
}
\end{table}

\subsection{Label generation}

To cater to diverse supervision requirements, we generate boundary and body labels based on ground truth data. A binary ground truth, denoted as $G_{final}$, is composed of foreground $G_{fg}$ and background $G_{bg}$. The initial step involves applying distance transformation to $G_{fg}$ to yield a distance map, representing the shortest distance from pixels in $G_{fg}$ to $G_{bg}$. In this distance map, distances equal to or less than a specified hyper-parameter are identified as part of the boundary, denoted as $G_{bound}$, while distances exceeding the hyper-parameter are assigned to the body, labeled as $G_{body}$. Notably, the labels for the remaining pixels in both $G_{bound}$ and $G_{body}$ are unequivocally designated as background. This process can be succinctly expressed as:
\begin{equation}
    G_{fg}^{j} \Rightarrow 
    \left\{
        \begin{matrix}
            G_{fg}^{j} \in {G_{{body}}}, \text{ }if\text{  }\text{  } \eta(G_{fg}^{j}, G_{bg}) > \alpha \\
            G_{fg}^{j} \in {G_{{bound}}},\ \text{ }if\text{  }\text{  } \eta(G_{fg}^{j}, G_{bg}) \le \alpha
        \end{matrix}
    \right.
\end{equation}
\begin{equation}
{{{G}}_{final}}={{G}_{bound}}\cup{{G}_{body}}
\end{equation}
where $\eta(G_{fg}^{j}, G_{bg})$ represents the distance from pixel $G_{fg}^{j}$ to $G_{bg}$, and the $G_{fg}^{j}$ represent pixels belonging to foreground. The hyper-parameter $\alpha$ is set to the default value of 1.

\subsection{Feature fusion and supervision block}
As depicted in Fig.\ref{fig2}, the designed FFS block primarily comprises of three parts: body and boundary pre-generation, feature fusion, and body and boundary supervision. For each FFS block, we leverage detailed and semantic information from the feature maps of both the encoder and decoder. To achieve comprehensive feature maps $F_{i}$, we concatenate the feature maps from the encoder with those from either the ASP\textunderscore OC block or the preceding FFS block. The encoder's feature maps contain extensive detailed information, whereas the decoder's feature maps from the preceding layer incorporate more semantic information. These contributions to the segmentation of boundaries and objects differ significantly. Therefore, we first process the concatenated feature map through two parallel convolutional branches, which pre-generate some boundary and body features. The feature map generated by two parallel convolutional layers can be represented as:
\begin{equation}
{{F}_{body,i}, {F}_{bound,i}}= Conv\left( {{F}_{i}} \right)
\end{equation}
where  $Conv$ represents the matrix of two 3×3 convolutions. ${{F}_{body,i}}$ and ${{F}_{bound,i}}$ denote pre-generation feature maps of body and boundary of the $i$-th layer, respectively. The next section will provide a detailed introduction to the feature fusion module.

The pre-generated feature maps undergo feature fusion and enhancement through a feature fusion module  $\Re\left(\bullet, \bullet\right)$. This process aims to integrate and enhance the boundary and body features previously generated by the convolutional branches. The $\Re\left(\bullet, \bullet\right)$ module performs the fusion by considering the relationship between body and boundary, which aims at enhancing segmentation quality. The process can be expressed as:
\begin{equation} 
( F _{body ,i}^{*}, F _{bound ,i}^{*}) =  \Re ( F _ { b o d y ,i }, F _ { b o u n d ,i } )
\end{equation}where $F _{body ,i}^{*}$ and $F _{bound ,i}^{*}$ represent the body and boundary feature maps after enhancement of the $i$-th layer, respectively. These enhanced feature maps are crucial in the training process as they are used for supervision. During the training phase, these feature maps are used to compute the loss, which is then optimized to enhance segmentation performance.

Finally, given two feature maps $F _{body ,i}^{*}$ and $F _{bound ,i}^{*}$, we propose to fuse them by a trainable parameter as the output of FFS. Specifically, the trainable parameter is continuously updated during the training process through backpropagation. Compared with previous works that directly use addition or concatenation for feature fusion\cite{7}\cite{8}, the proposed FFS introduces the interaction between the features from the boundary and body in training. Fusion by trainable parameter can be expressed as :
\begin{equation}
{{F}_{out,i}}=\lambda F_{body,i}^{*}+ F_{bound,i}^{*}
\end{equation}
where ${{F}_{out,i}}$ represents the output feature map of FFS of the $i$-th layer. In particular, the output of the final FFS module is denoted as $\hat{F} $. $\lambda $ is a trainable parameter, and the initial value is set to 1.

\begin{figure}[!t]
\centerline{\includegraphics[width=0.5\columnwidth]{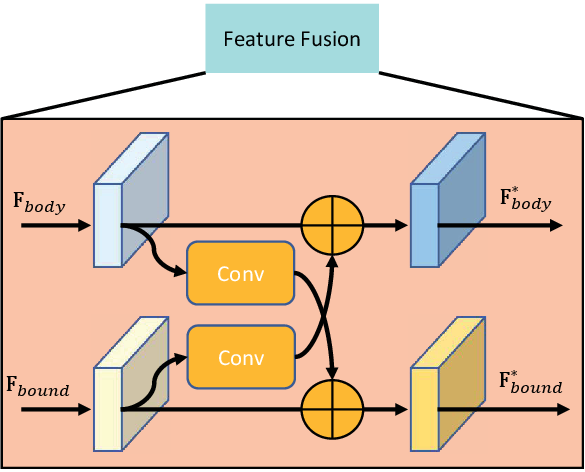}}
\caption{Details of boundary and body feature fusion block. Conv means successive convolution operations.}
\label{fig3}
\end{figure}

\subsection{Feature fusion module}
Here We give a detailed introduction of the feature fusion module (FFM)$\Re\left(\bullet, \bullet\right)$ in FFS. Due to the close relationship between semantic segmentation and boundary segmentation tasks, we introduce a novel feature fusion block. This block facilitates bidirectional information flow between the body and boundary streams. FFM is illustrated in Fig. \ref{fig3}. It receives two input feature maps $F_{body}$ and $F_{bound}$ from parallel convolution layers. The $F_{body}$ and $F_{bound}$ are forwarded to a convolution layer, respectively. Those features are subsequently added back to the original features $F_{body}$ and $F_{bound}$ to derive the complementary features $F_{body}^{*}$ and $F_{bound}^{*} $. The process in the feature fusion block is as follows:
\begin{equation}
{F}_{body}^{*}={{F}_{body}}+Conv({{F}_{bound}})
\end{equation}
\begin{equation}
{F}_{bound}^{*}={{F}_{bound}}+Conv({{F}_{body}})
\end{equation}
where $Conv$ means a convolution layer. 

\subsection{Loss function}
As shown in Fig. \ref{fig2}, there are multiple tasks for the proposed network, including body map $F_{body}^{*}$, boundary map $F_{bound}^{*} $, and final segmentation map $\hat{F} $. Specially, we adopt auxiliary supervised losses for $F_{body}^{*}$ and $F_{bound}^{*} $, receptively. The total loss L is computed as:
\begin{equation}
\begin{split}
L = &L_{seg} ( \hat{F}, G_{final} ) +\sum_{i}( L_{body} ( F_{body,i}^{*},G_{body}) 
+ \\& L_{bound}(F_{bound,i}^{*},G_{bound}))
\end{split}
\end{equation}
where $L_{seg}$, $L_{body}$, and $L_{bound}$ represent the loss for the final segmentation map, body map, and boundary map, respectively.

For simplicity in optimization, we employ the same loss function across all training stages. The loss $L_{seg}$, $L_{body}$ and $L_{bound}$ are given by two terms: $L_{wbce}$ is the weight binary cross-entropy loss, while represents Dice loss. The weight of $L_{dice}$ is inversely proportional to the ratio of foreground to background pixels in the label, aiming to prioritize attention on the foreground region within the network. The formulations are shown as follows:
\begin{equation}
L ^ { * } ( P , G ) = \lambda _ { 1 } L _ { w b c e } ( P , G ) + \lambda _ { 2 } L _ { d i c e } ( P , G )
\end{equation}
\begin{equation}
L _ { w b c e } (G, P) = -\sum_{i} w_i \left(G_i \cdot \log(P_i) + (1 - G_i) \cdot \log(1 - P_i)\right)
\end{equation}
\begin{equation}
 L _ { d i c e }  = 1 - \frac{2|G\cap P|}{|G| + |P|}
\end{equation}
where $\lambda _ { 1 }$ and $\lambda _ { 2 }$ are two hyper-parameters that control the weight among the two losses and we set them 1 and 10, respectively, as default. $L ^ { * }$ represents $L_{seg}$, $L_{body}$ and $L_{bound}$. $P$ is the prediction map, $G$ is the mask. The definition of the weight $w_i$ is as follows:
\begin{equation}
w_i = (log(\beta + \frac{G_i}{\sum G_i}))^{-1}
\end{equation}
where $\beta$ is a hyperparameter, and we set 1 as the default value.

\section{Experiments}
This section describes the experimental settings and datasets used for ultrasound image segmentation to evaluate and compare the proposed method.
\subsection{Experiment settings}
The experiments were conducted using PyTorch 1.9.1 on Ubuntu 18.04.5, with training performed on a single RTX 1080Ti GPU. Optimization was carried out using the Adam algorithm, starting with an initial learning rate of 0.001. we implemented a polynomial learning rate policy with a power of 0.9 to adjust the learning rate. Batch sizes were set to 2, 4, and 6, and maximum epochs were configured to 300, 100, and 350 for datasets BUSI, UNS, and UHES, respectively. Online augmentation techniques included random resizing with a scale of 0.75 to 1.5, random cropping, and random horizontal flips.

\subsection{Dataset}
To validate the effectiveness of our proposed method, we selected three ultrasound image datasets featuring various lesions and tissues for comprehensive  experiments, including segmentation tasks for breast cancer, nerve, and hemangioma. we 
\begin{enumerate}
\item BUSI \cite{35}: The BUSI dataset, constituting a repository of breast cancer data, comprises 780 images obtained from 600 female patients, with an average image resolution of 500×500. Image acquisition was conducted using the LOGIQ E9 US system and LOGIQ E9 Agile US. The BUSI dataset includes 133 instances of normal breast masses, 437 benign cases, and 210 malignant occurrences. In accordance with the configuration specified in the UNeXt \cite{33}, both the training and test sets were resized to dimensions of 512×512 for consistency during both phases.
\item UNS \cite{36} :  The ultrasound nerve segmentation (UNS) Kaggle challenge in 2016 involved identifying and marking the brachial plexus (BP) nerves in ultrasound images. This was a difficult task because many of the training images did not contain the BP area. The challenge used a dataset of 5,635 training images and 5,508 test images with a size of 580×420 pixels. 
\item UHES: The ultrasound fetal hemangioma segmentation (UHES) dataset was graciously made available by the Children's Hospital of Chongqing Medical University, comprising a total of 215 annotated images. In light of the practicalities associated with image processing and the average dimensions of the dataset images, a judicious decision was made to resize each image to a standardized resolution of 448×256. This resizing strategy was implemented to enhance the uniformity and facilitate the subsequent analyses of the dataset.
\end{enumerate}

\begin{table*}[!ht]
\caption{The segmentation results of different CNN-based networks on BUSI, UNS, and UHES datasets are provided, highlighting the best-performing results in bold. Mean and standard deviation values are included. * mean the official result}
\centering
\begin{tabular}{c c c c c}
\hline
Method & Metric & BUSI & UNS & UHES \\
\hline
\multirow{2}{*}{U-Net{\cite{7}}} & DSC(\%) & 65.19$\pm$7.19 & 70.86$\pm$10.45 & 78.24$\pm$24.00 \\
& HD(mm) & 9.93$\pm$0.17 & 3.88$\pm$1.25 & 9.06$\pm$0.51 \\
\hline
\multirow{2}{*}{DeepLabV3+{\cite{37}}} & DSC(\%) & 77.76$\pm$8.92 & 69.61$\pm$12.32 & 85.02$\pm$9.69 \\
& HD(mm) & 7.66$\pm$0.27 & 3.98$\pm$1.56 & 8.30$\pm$0.30 \\
\hline
\multirow{2}{*}{LinkNet{\cite{8}}} & DSC(\%) & 72.70$\pm$9.77 & 74.68$\pm$11.46 & 82.97$\pm$6.19 \\
& HD(mm) & 8.41$\pm$0.70 & 3.61$\pm$0.95 & 8.55$\pm$0.22 \\
\hline
\multirow{2}{*}{DBBS-Net{\cite{38}}} & DSC(\%) & 79.34$\pm$9.43 & 75.69$\pm$8.65 & 84.85$\pm$10.49 \\
& HD(mm) & 8.05$\pm$0.32 & 3.55$\pm$0.85 & 9.84$\pm$0.37 \\
\hline
\multirow{2}{*}{UNeXt{\cite{39}}} & DSC(\%) & 78.17$\pm$2.57 & 69.24$\pm$9.92 & 83.11$\pm$5.74 \\
& HD(mm) & 8.16$\pm$0.39 & 4.06$\pm$1.95 & 8.57$\pm$0.27 \\
\hline
\multirow{2}{*}{MSSA-Net$^*${\cite{47}}} & DSC(\%) & 80.65 & - & - \\
& HD(mm) & - & - & - \\
\hline
\multirow{2}{*}{V2-CE-CD$^*${\cite{48}}} & DSC(\%) & 80.23 & - & - \\
& HD(mm) & - & - & - \\
\hline
\multirow{2}{*}{HEAT-Net$^*${\cite{49}}} & DSC(\%) & 74.1  & - & - \\
& HD(mm) & - & - & - \\
\hline
\multirow{2}{*}{EHA-Net$^*${\cite{50}}} & DSC(\%) & 80.64  & - & - \\
& HD(mm) & - & - & - \\
\hline
\multirow{2}{*}{DSEU-net$^*${\cite{51}}} & DSC(\%) & 78.51 ± 1.87  & - & - \\
& HD(mm) & - & - & - \\
\hline
\multirow{2}{*}{NU-net$^*${\cite{52}}} & DSC(\%) & 78.62 ± 1.38  & - & - \\
& HD(mm) & - & - & - \\
\hline
\multirow{2}{*}{DBF-Net(Ours)} & DSC(\%) & \textbf{81.05$\pm$10.44} & \textbf{76.41$\pm$5.52} & \textbf{87.75$\pm$4.18} \\
& HD(mm) & \textbf{7.35$\pm$0.27} & \textbf{3.52$\pm$0.52} & \textbf{7.35$\pm$0.27} \\
\hline
\end{tabular}
\end{table*}

\begin{figure}[!t]
\centerline{\includegraphics[width=\linewidth]{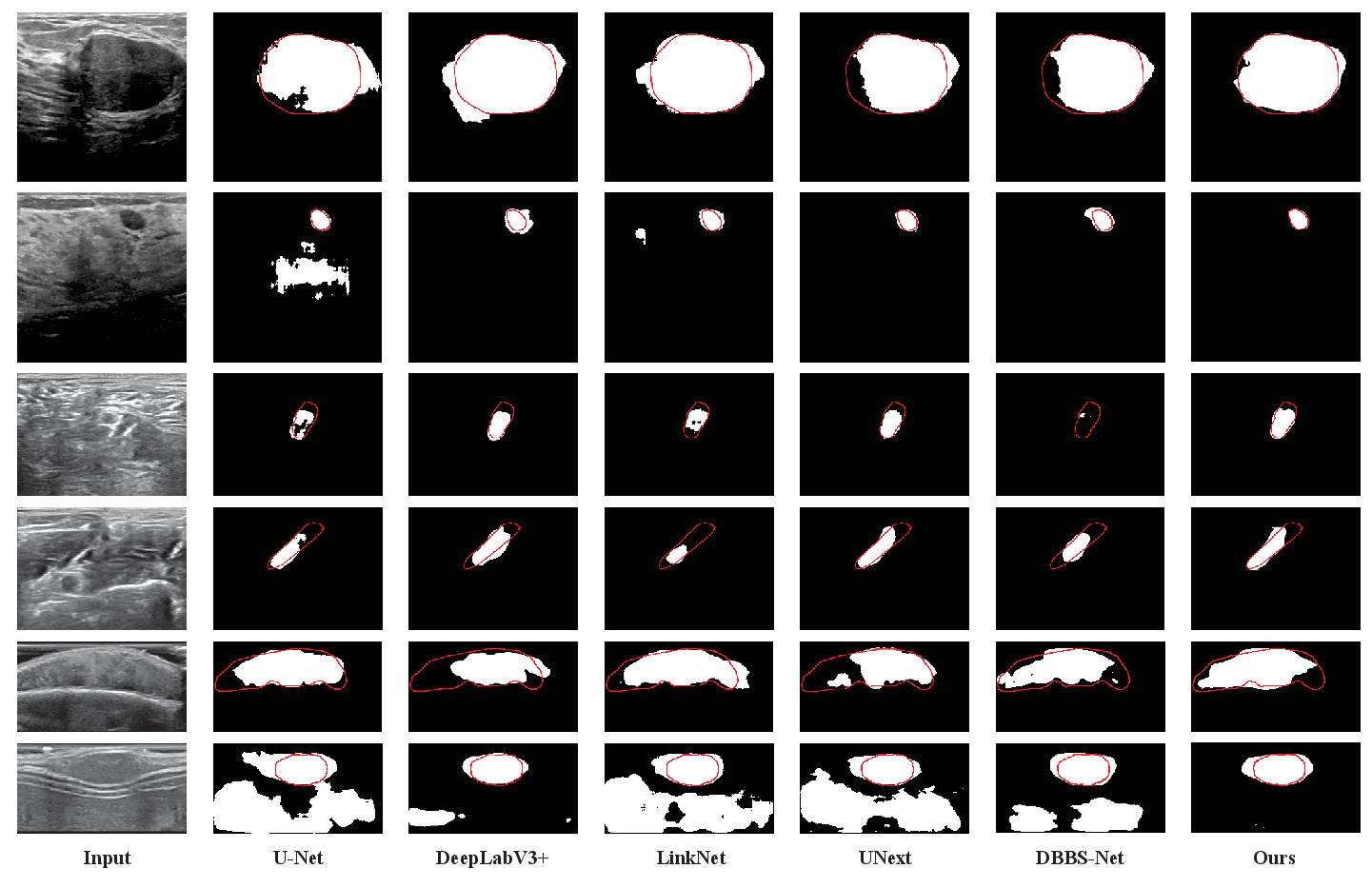}}
\caption{Comparison of qualitative results between U-Net, DeepLabV3+, LinkNet, UNeXt, DBBS-Net and the proposed method for breast cancer
segmentation using BUSI dataset, brachial plexus nerves segmentation using UNS, infantile hemangioma segmentation using UHES. The red curve outlines the boundaries of the lesions.
The first and second rows show results from the BUSI dataset, while the third and fourth rows depict results from UNS, and the fifth and sixth rows present results from UHES.}
\label{fig4}
\end{figure}

\begin{figure*}[!htp]
	\centering
	\begin{minipage}{0.32\linewidth}
		\centering
		\includegraphics[width=1\linewidth]{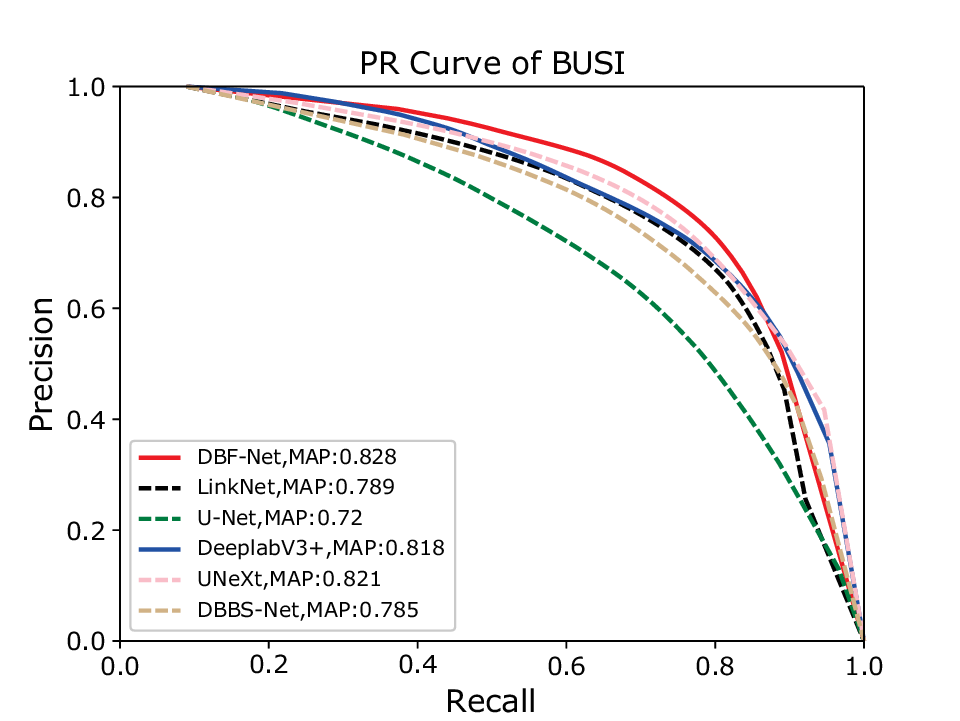}
	\end{minipage}
	\begin{minipage}{0.32\linewidth}
		\centering
		\includegraphics[width=1\linewidth]{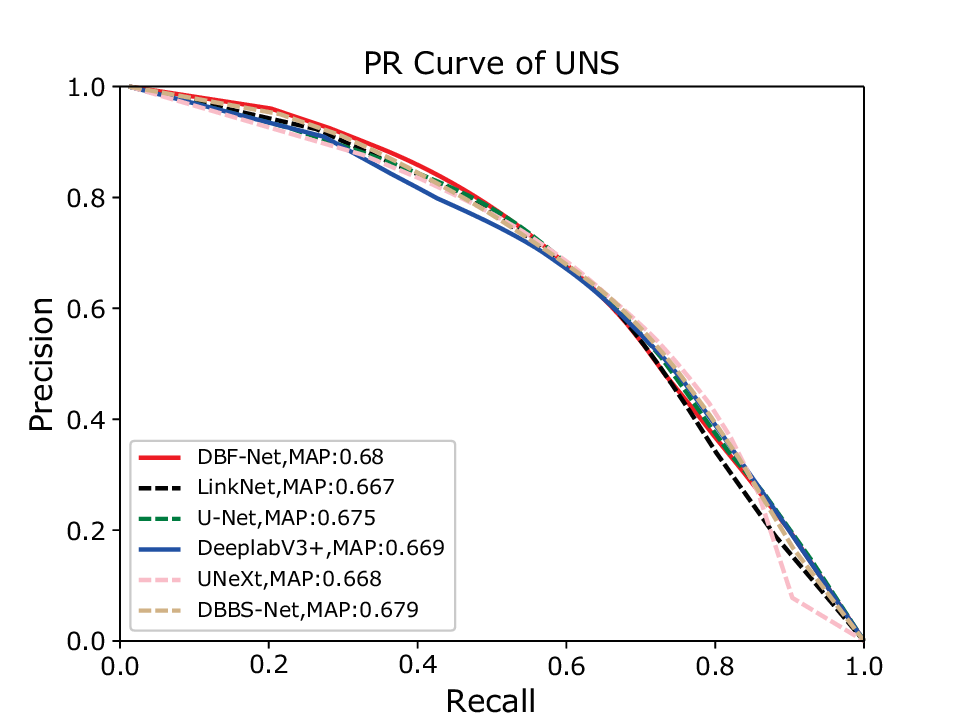}
	\end{minipage}
	\begin{minipage}{0.32\linewidth}
		\centering
		\includegraphics[width=1\linewidth]{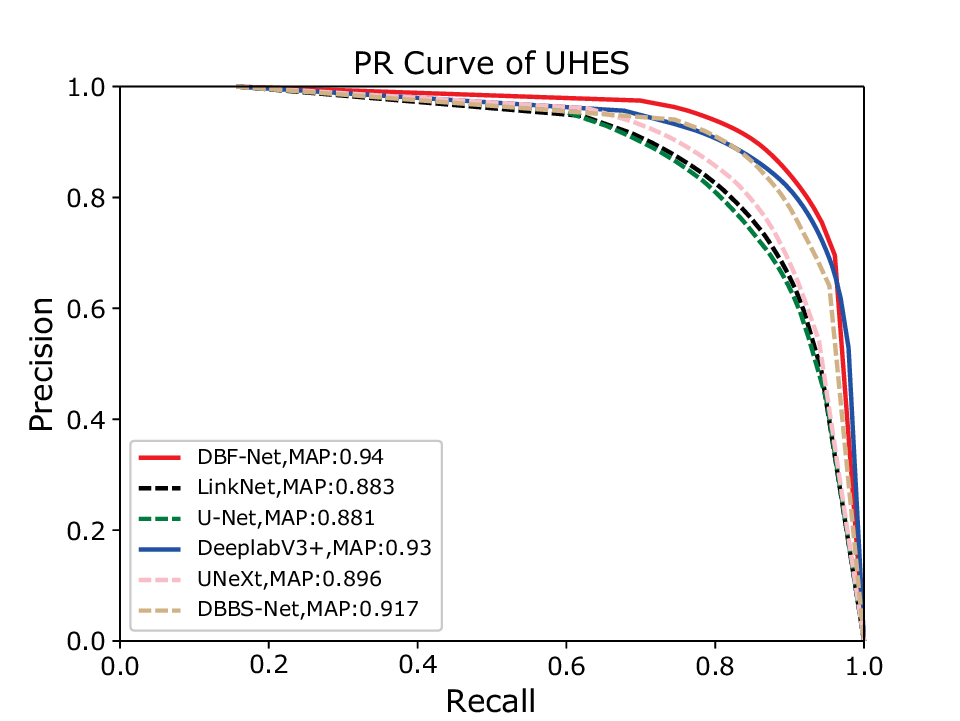}
	\end{minipage}
	%\qquad
	\begin{minipage}{0.32\linewidth}
		\centering
		\includegraphics[width=1\linewidth]{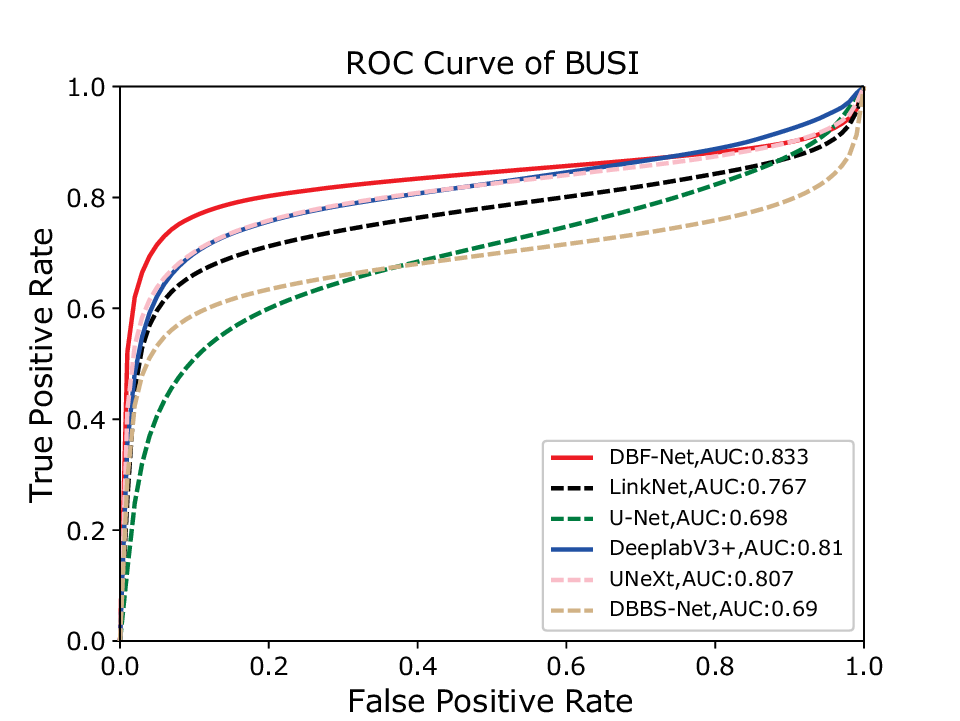}
	\end{minipage}
	\begin{minipage}{0.32\linewidth}
		\centering
		\includegraphics[width=1\linewidth]{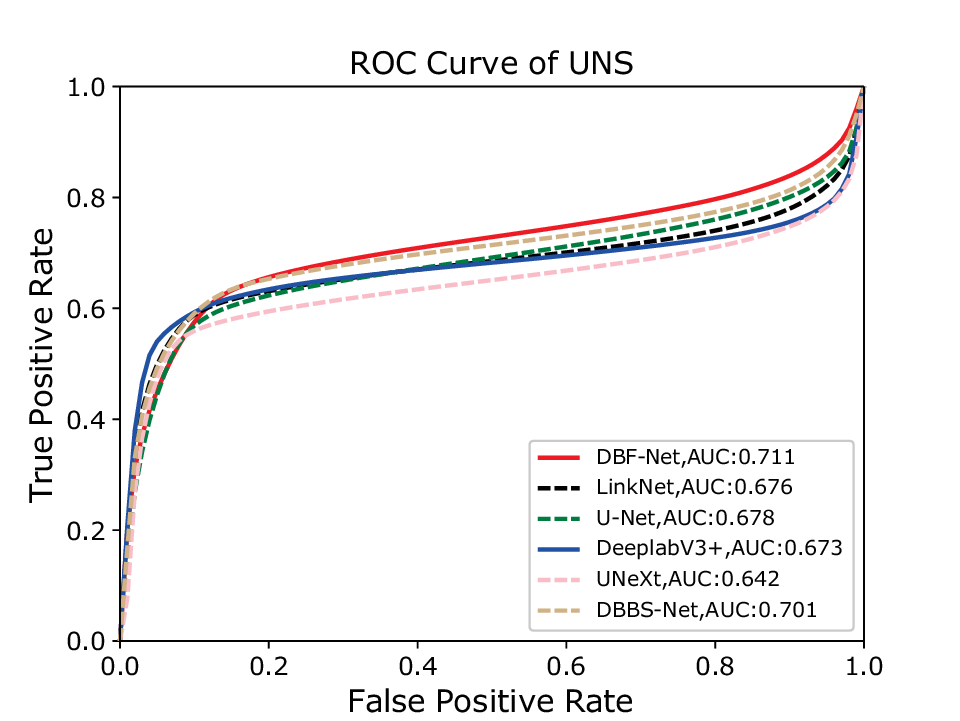}
	\end{minipage}
	\begin{minipage}{0.32\linewidth}
		\centering
		\includegraphics[width=1\linewidth]{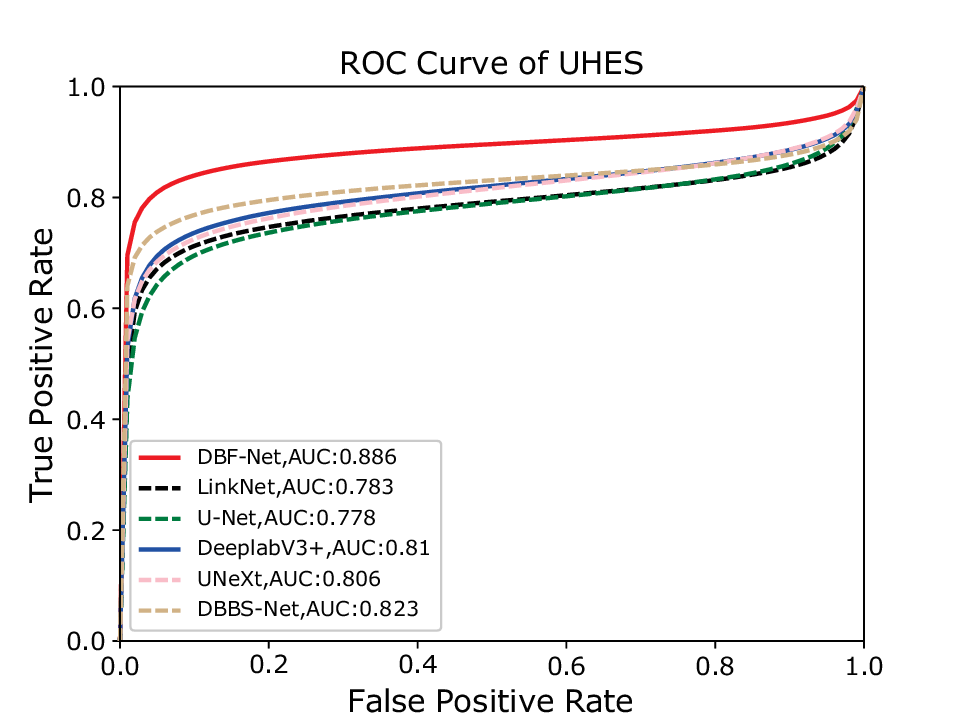}
	\end{minipage}
\caption{P-R and ROC curves of DBF-Net, LinkNet, U-Net, DeepLabV3+, UNeXt, and DBBS-Net on BUSI, UNS, and UHES.}
\label{fig5}
\end{figure*}

\subsection{Comparison with CNN-based methods}
To quantitatively assess the segmentation performance of various methods in ultrasound image analysis, two widely employed segmentation metrics are utilized: the Dice Similarity Coefficient (DSC) and the Hausdorff distance (HD). To ensure fairness, all experiments are conducted using the same pipeline. We employ five-fold cross-validation to generalize the models' performance. To evaluate the robustness and efficacy of the proposed method in this paper, we initially conducted a comparative analysis against state-of-the-art convolutional methods. The comparative methods include U-Net\cite{7}, DeepLabV3+\cite{37}, LinkNet \cite{8}, UNeXt \cite{39}, and DBBS-Net \cite{38}. In addition, we also present some networks specifically designed for segmenting ultrasound images, such as MSSA-Net\cite{47}, V2-CE-CD\cite{48}, HEAT-Net\cite{49}, EHA-Net\cite{50}, DSEU-net\cite{51}, and NU-net\cite{52}. The average performance of these models is presented in Table 2. From Table 2, it is evident that that DBF-Net achieves superior DSC of 81.05$\pm$10.44\%, 76.41$\pm$5.52 and 87.75$\pm$4.18 on breast cancer, hemangioma and brachial plexus nerves segmentation, respectively. The HD values of DBF-Net on BUSI, UNS, and UHES are 7.35$\pm$0.27, 3.52$\pm$0.52, and 7.35$\pm$0.27, respectively. Fig. \ref{fig4} presents the qualitative results of our model and state-of-the-art methods on three datasets. In comparison to alternative techniques, the segmentation outcomes of DBF-Net, our proposed approach, closely resemble the ground truth.

\begin{figure}[!htp]
\centerline{\includegraphics[width=\linewidth]{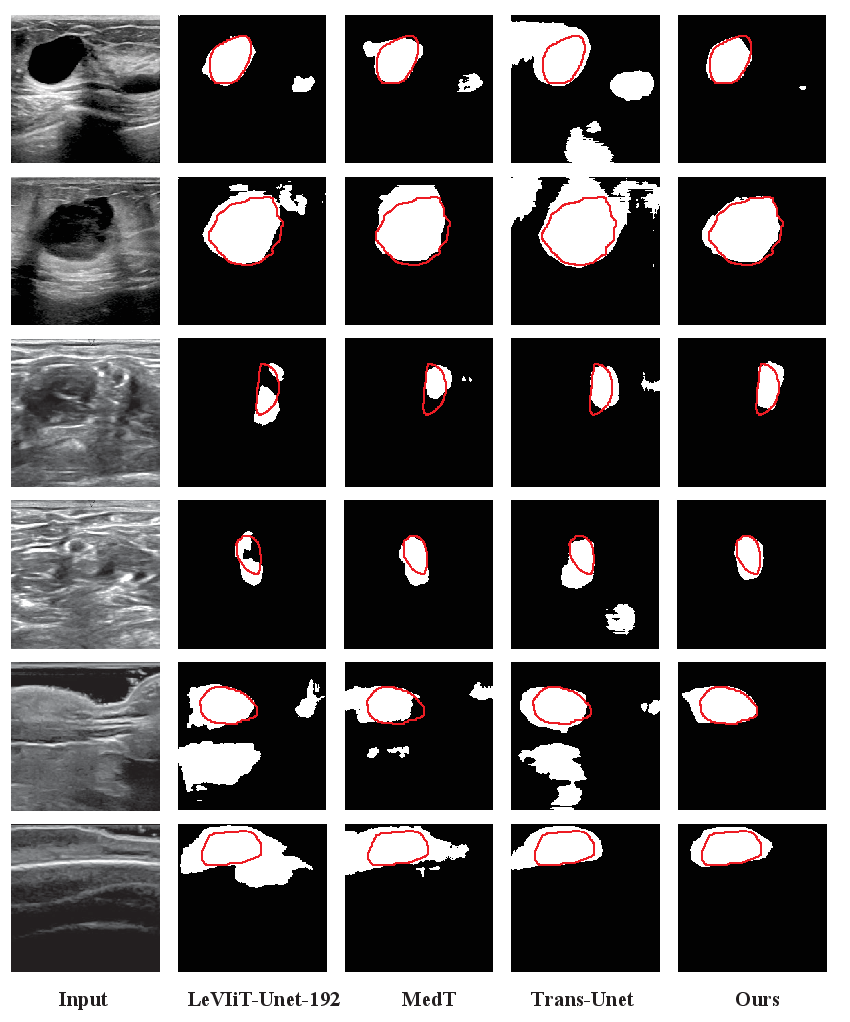}}
\caption{Comparison of qualitative results between LeViT-UNet-192,
MedT, TransUnet and the proposed method for breast cancer segmentation
using BUSI dataset, brachial plexus nerves segmentation using
UNS, infantile hemangioma segmentation using UHES. The red curve represents the boundary of the breast tumor. The first and second rows show the results for the BUSI dataset, the third and fourth rows depict the results for UNS, and the fifth and sixth rows present the results for UHES.}
\label{fig6}
\end{figure}

\begin{figure*}[htbp]
	\centering
	\begin{minipage}{0.32\linewidth}
		\centering
		\includegraphics[width=1\linewidth]{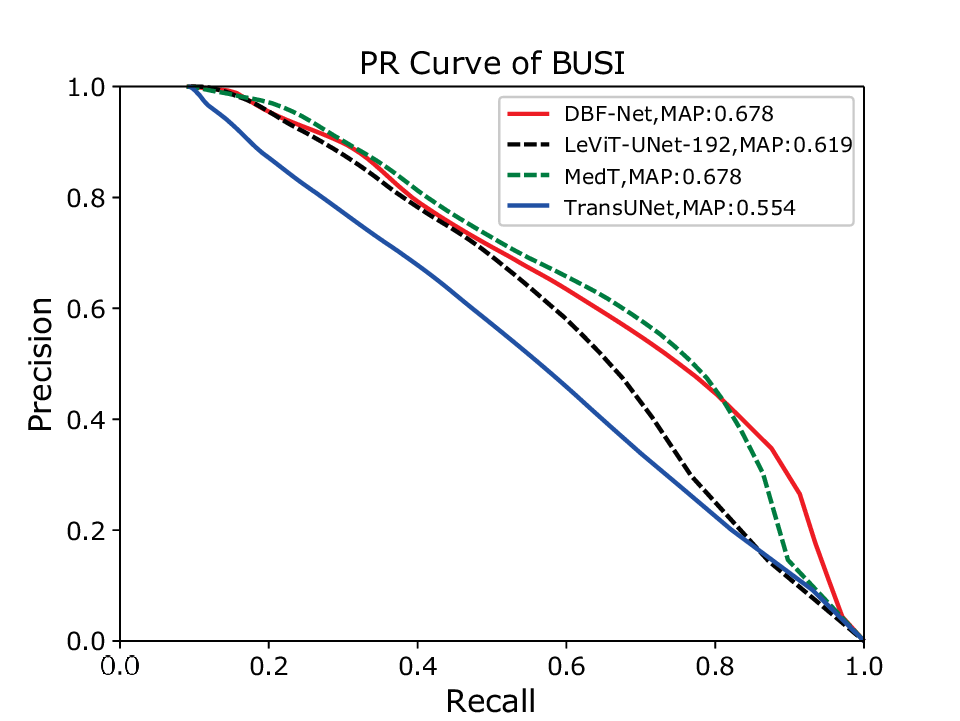}
	\end{minipage}
	\begin{minipage}{0.32\linewidth}
		\centering
		\includegraphics[width=1\linewidth]{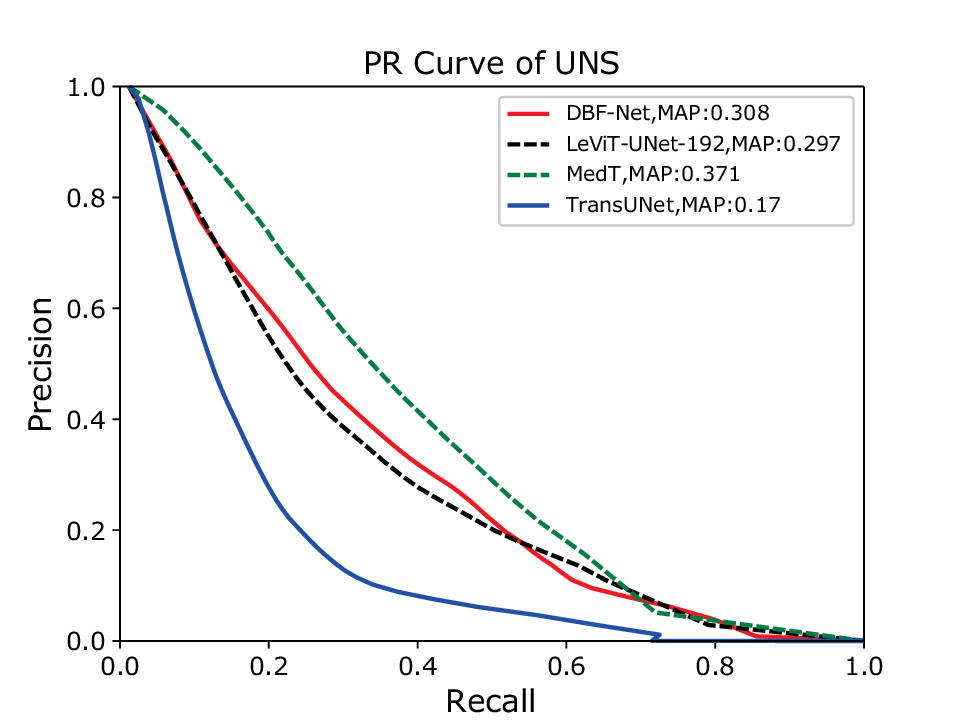}
	\end{minipage}
	\begin{minipage}{0.32\linewidth}
		\centering
		\includegraphics[width=1\linewidth]{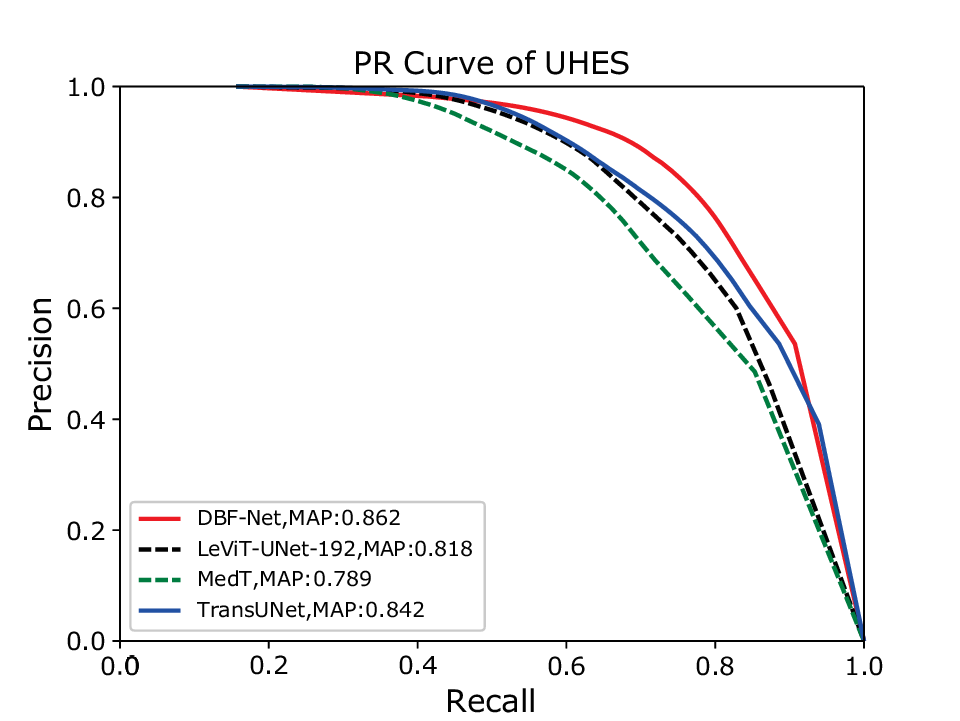}
	\end{minipage}
	%\qquad
	\begin{minipage}{0.32\linewidth}
		\centering
		\includegraphics[width=1\linewidth]{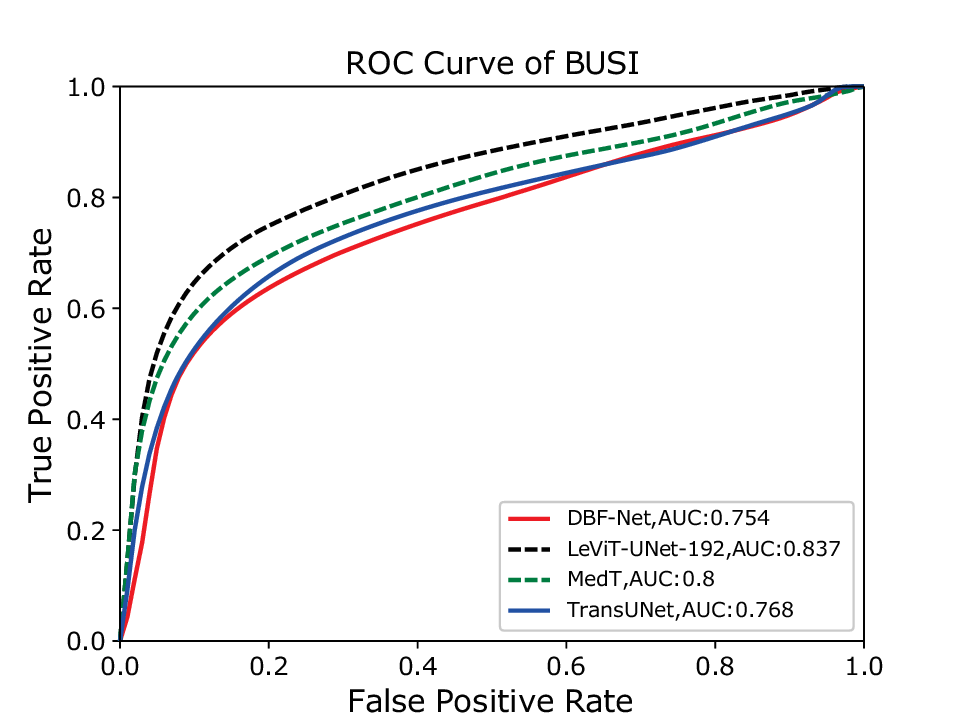}
	\end{minipage}
	\begin{minipage}{0.32\linewidth}
		\centering
		\includegraphics[width=1\linewidth]{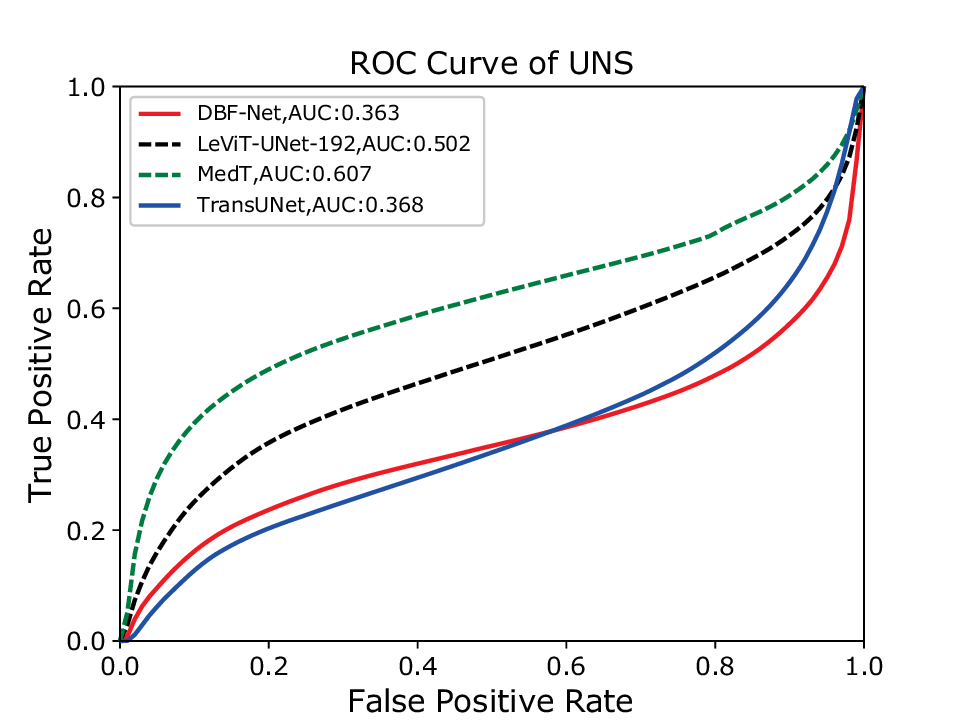}
	\end{minipage}
	\begin{minipage}{0.32\linewidth}
		\centering
		\includegraphics[width=1\linewidth]{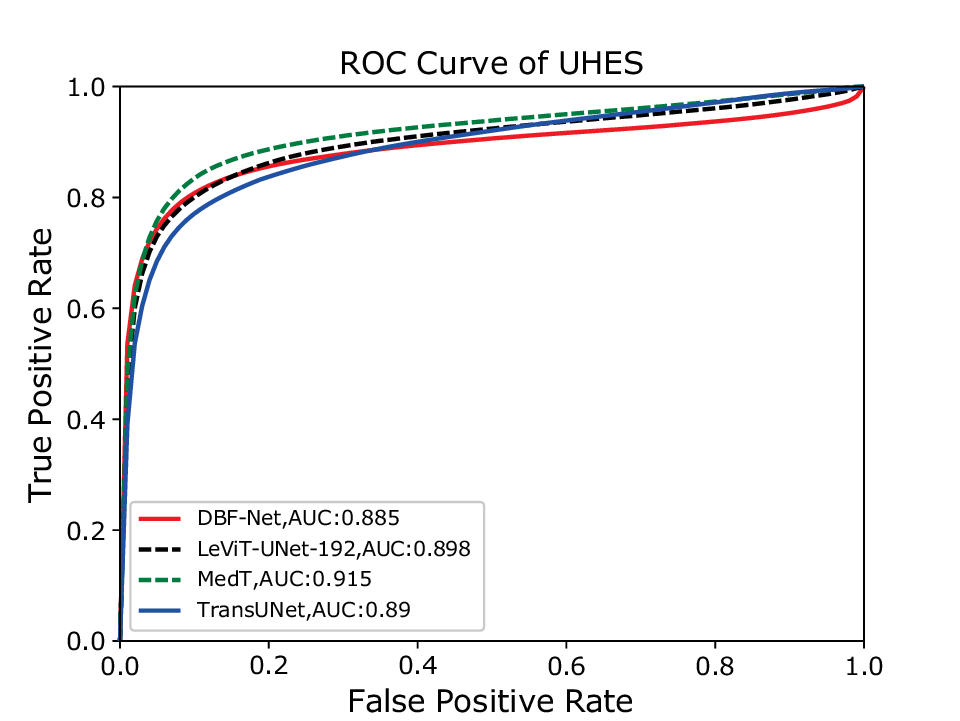}
	\end{minipage}
\caption{P-R and ROC curves of DBF-Net, LeViT-UNet-192, MedT and
TransUNet on BUSI, UNS and UHES.}
\label{fig7}
\end{figure*}

Fig. \ref{fig5} displays the precision-recall (P-R) and Receiver Operating Characteristic (ROC) curves for several CNN-based methods on BUSI, UNS, and UHES, along with the AUC
and MAP scores on the corresponding curves. Our method outperforms other approaches by  attaining the highest AUC and MAP values on BUSI, UNS, and UHES datasets. Comparison of the precision-recall (P-R) and Receiver Operating Characteristic (ROC) curves indicates that DBF-Net  achieves the highest confidence levels in segmenting BUSI, UNS, and UHES.

\begin{table*}[!ht]
\caption{This graph shows the segmentation outcomes of various transformer networks on BUSI, UNS, and UHES with the best results highlighted in bold text. Mean and standard deviation are also included.}
\centering
\begin{tabular}{c c c c c c}
\hline
 Method  & Metric & BUSI  & UNS    & UHES   \\
 \hline
\multicolumn{1}{l}{\multirow{2}{*}{LeViT-UNet-192{\cite{40}}}}
& DSC(\%) & 44.37$\pm$13.31  & 40.07$\pm$16.85  &  64.01$\pm$15.09\\
& HD(mm) & 8.71$\pm$0.32  & 4.25$\pm$0.78 & 8.00$\pm$0.15   \\
\hline
\multirow{2}{*}{MedT{\cite{41}}}                      
 & DSC(\%) & 52.27$\pm$11.1  & 37.69$\pm$17.84  & 63.15$\pm$14.25 \\
& HD(mm) & 7.95$\pm$0.31  & 4.84$\pm$0.64 & 8.92$\pm$0.08\\
\hline                    
\multirow{2}{*}{TransUNet{\cite{42}}}                 & DSC(\%) & 31.40$\pm$8.95 & 36.68$\pm$15.23  & 69.20$\pm$4.09  \\
& HD(mm) & 10.03$\pm$0.05   & 4.51$\pm$0.84 & 7.98$\pm$0.26  \\
\hline                     
\multirow{2}{*}{DBF-Net(Ours)} 
& DSC(\%) & \textbf{55.49$\pm$2.12} & \textbf{56.49$\pm$10.12} & \textbf{70.20$\pm$4.09} \\
 & HD(mm) & \textbf{6.67$\pm$0.04 } & \textbf{2.76$\pm$0.34 } & \textbf{7.63$\pm$0.16 } \\
\hline
\end{tabular}
\end{table*}

% \begin{table*}[!ht]
% \caption{The segmentation results of different transformer networks on BUSI, UNS, and UHES. The best results are marked with bold text. Both mean and standard deviation are given.}
% \centering
% \begin{tabular}{c c c c c c}
% \hline
%  Method  & Resolution & Metric & BUSI  & UNS    & UHES   \\
%  \hline
% \multicolumn{1}{l}{\multirow{2}{*}{LeViT-UNet-192{\cite{40}}}} & \multirow{2}{*}{224×224} & DSC(\%) & 44.37$\pm$13.31  & 40.07$\pm$16.85  &  64.01$\pm$15.09\\
% \multicolumn{1}{l}{} &            & HD(mm) & 8.71$\pm$0.32  & 4.25$\pm$0.78 & 8.00$\pm$0.15   \\
% \hline
% \multirow{2}{*}{MedT{\cite{41}}}                              
% & \multirow{2}{*}{224×224} & DSC(\%) & 52.27$\pm$11.1  & 37.69$\pm$17.84  & 63.15$\pm$14.25 \\
%  &      & HD(mm) & 7.95$\pm$0.31  & 4.84$\pm$0.64 & 8.92$\pm$0.08\\
% \hline                    
% \multirow{2}{*}{TransUNet{\cite{42}}}                          
% & \multirow{2}{*}{224×224} & DSC(\%) & 31.40$\pm$8.95 & 36.68$\pm$15.23  & 69.20$\pm$4.09  \\
%   &    & HD(mm) & 10.03$\pm$0.05   & 4.51$\pm$0.84 & 7.98$\pm$0.26  \\
% \hline                     
% \multirow{2}{*}{DBF-Net(Ours)} 
% & \multirow{2}{*}{224×224} & DSC(\%) & \textbf{55.49$\pm$2.12} & \textbf{56.49$\pm$10.12} & \textbf{70.20$\pm$4.09} \\
%  &     & HD(mm) & \textbf{6.67$\pm$0.04 } & \textbf{2.76$\pm$0.34 } & \textbf{7.63$\pm$0.16 } \\
% \hline
% \end{tabular}
% \end{table*}

\subsection{Comparison with transformer-based methods}
We also compared DBF-Net with some transformer-based models, like LeViT-UNet-192 \cite{40}, MedT \cite{41} and TransUNet\cite{42}. It should be noted that we adjusted the input resolution
of images to 224×224 due to the GPU's limited memory.Table 3 presents the quantitative evaluation results of different segmentation methods. From Table 3, it is clear that DBF-Net consistently outperforms the compared transformer-based methods across both Dice Similarity Coefficient (DSC) and Hausdorff Distance (HD) metrics. In Fig. \ref{fig6}, we provide qualitative results of DBF-Net and transformer-based networks on BUSI, UNS, and UHES. Comparing our proposed method to other segmentation methods, we find it to be more effective visually. In Fig.\ref{fig7}, we draw the PR curves and ROC curves of LeViTUNet-192, MedT, Trans-Unet and the proposed method on three datasets. According to the P-R curves, the proposed method outperformed LeViT-UNet-192, MedT, and Trans-Unet on UHES and ranked second on BUSI and UNS. However, the ROC curves indicate that the proposed method performs suboptimally.

\begin{table}[!h]
\caption{Comparison of a number of parameters.}
\centering
\scalebox{0.8}{
\begin{tabular}{c c c c c c}
\hline
Method   & U-Net   & DeepLabV3+ & LinkNet    & DBBS-Net & UNeXt \\
\ Params(M) & 17.8  & 5.8  & 11.5   & 2.7  & 1.47  \\
\hline
Method   & LeViT-UNet-192 & MedT    & Trans-Unet & -   & Ours  \\
\ Params(M) & 15.9           & 1.6        & 91.7       & -       & 3.2 \\
\hline
\end{tabular}
}
\end{table}

\begin{table}[!ht]
  \centering
  \caption{Ablation experiments for body and boundary supervision
block and feature fusion strategy. baseline means keeping
the encoder and ASP\_OC block in DBF-Net and using
output stride 16 to generate the segmentation map
}
\begin{tabular}{ccp{4.19em}p{6.5em}cc}
\hline
\multirow{2}{*}{} 
& \multicolumn{3}{c}{FFS number and setting} 
& \multicolumn{1}{c}{\multirow{2}{*}{Dice(\%)}} 
& \multicolumn{1}{c}{\multirow{2}{*}{HD(mm)}} \\
\cline{2-4}   & \multicolumn{1}{c}{1}     
& \multicolumn{1}{c}{2} & Feature fusion \\
\hline
\multicolumn{1}{p{4.19em}}{Baseline} 
&    & \multicolumn{1}{c}{} & \multicolumn{1}{c}{} & 79.91$\pm$9.43 & 7.39$\pm$0.35 \\
\hline
\multicolumn{1}{c}{\multirow{4}{*}{+FFS}} 
& \multicolumn{1}{c}{\checkmark} & \multicolumn{1}{c}{} & \multicolumn{1}{c}{} 
& 79.85$\pm$10.03   & 7.41$\pm$0.29 \\
&       & \multicolumn{1}{c}{\checkmark}     & \multicolumn{1}{c}{} 
& 80.73$\pm$7.68 & 7.37$\pm$0.22  \\
\cline{2-6}          & \multicolumn{1}{c}{\checkmark} & \multicolumn{1}{c}{} & \multicolumn{1}{c}{\checkmark}     
& 80.34$\pm$10.75 & 7.38$\pm$0.21 \\
 &       & \multicolumn{1}{c}{\checkmark}     & \multicolumn{1}{c}{\checkmark}     
    & 81.05$\pm$10.44 & 7.35$\pm$0.27 \\
    \hline
    \end{tabular}%
\end{table}%

\subsection{Complexity analysis}
In Table 4, the number of parameters for the proposed method and other networks compared are tabulated.Comparing the proposed method with U-Net, we find that it has approximately three times fewer parameters. Moreover, even when compared to other lightweight networks like UNeXt, our model exhibits a competitive parameter count. Our findings demonstrate that our proposed method achieves an optimal balance between model complexity and segmentation accuracy.

\subsection{Ablation study}
\begin{figure}[!ht]
\centerline{\includegraphics[width=\columnwidth]{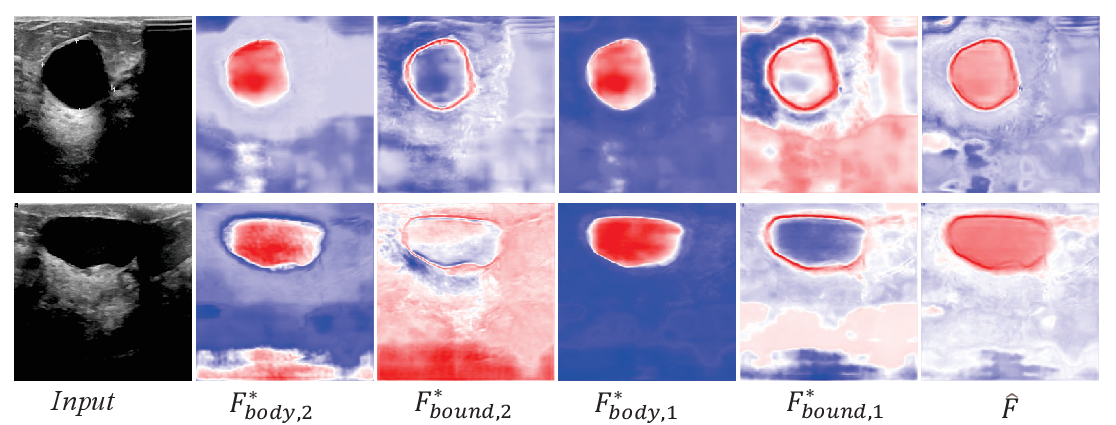}}
\caption{Example output of feature fusion module and final output for ultrasound image segmentation.}
\label{fig8}
\end{figure}

Table 5 shows ablation studies on body and boundary supervision block and feature fusion strategy. We perform all experiments on the BUSI dataset. From Table 5, we can note the following: (I) The proposed FFS module enhances the network’s performance over the baseline.(II) Furthermore, we obtain additional performance improvement by incorporating
the feature fusion module into the FFS module. Results presented in Table 5 indicate that the optimal configuration for ultrasound image segmentation is achieved with two FFS modules combined with feature fusion
operations as illustrated in Fig. \ref{fig2}. As a result of these findings, it appears that the FFS module, specifically when paired with feature fusion, can significantly enhance the segmentation performance of neural networks. Qualitative results of the feature fusion and final output can be seen in Fig. \ref{fig8}. We can see that $F_{body,1}^{*}$ and $F_{bound,1}^{*}$ are complementary to each other.

\begin{table}[!h]
\caption{Ablation experiments on body and boundary supervision. }
\centering
\begin{tabular}{cccccc}
\hline
Method & \multicolumn{1}{l}{$L_{seg}$} & \multicolumn{1}{l}{$L_{body}$} & \multicolumn{1}{l}{$L_{bound}$} & DSC(\%) & HD(mm) \\
\hline
\multirow{4}{*}{DBF-Net} & \checkmark &   &   & 78.79$\pm$11.65 & 7.59$\pm$0.30  \\
                          & \checkmark & \checkmark &   & 79.91$\pm$9.43 & 7.39$\pm$0.35 \\
                          & \checkmark &   & \checkmark & 79.79$\pm$9.93 & 7.67$\pm$0.27 \\
                          & \checkmark & \checkmark & \checkmark & 81.05$\pm$10.44 & 7.35$\pm$0.27 \\
\hline
\end{tabular}
\end{table}

The DBF-Net architecture integrates body and boundary supervision as a vital component. In this research, we assess the efficacy of this supervision scheme and provide the associated quantitative outcomes in Table 6.Our findings reveal that the incorporation of additional loss functions, specifically $L_{body} $  and $L_{bound}$, can effectively enhance the segmentation performance, with a more pronounced effect observed when both are utilized in combination. 

\begin{table}[!h]
  \centering
  \caption{Ablation experiments for the parameter $\lambda$ in Equation (5).}
  \begin{tabular}{lccc}
    \hline
    Method &Cross-validation &\textit{$\lambda$ in FFS-1} & \textit{$\lambda$ in FFS-2} \\
    \hline
    w/ FFM & 1& 0.94 & 1.05 \\
           & 2& 1.01 & 1.06 \\
           & 3 & 1.10 & 1.04 \\
           & 4 & 1.14 & 1.01 \\
           & 5 & 1.18 & 0.99 \\
    \hline
    w/o FFM & 1  & 1.21 & 1.38 \\
            & 2  & 1.40 & 1.40 \\
            & 3  & 1.49 & 1.45 \\
            & 4  & 1.55 & 1.43 \\
            & 5  & 1.61 & 1.32 \\
    \hline
  \end{tabular}
\end{table}

In Table 7, we present the conclusive outcomes derived from the training of the trainable parameter $\lambda$ on the BUSI dataset. Through a meticulous examination of the fluctuations exhibited by $\lambda$ under conditions involving both the inclusion and exclusion of the Feature Fusion and Supervision block (FFM), a discernible pattern emerges. Our scrutiny reveals that upon the incorporation of FFM, the disparities in weights between the boundary and the main body components are rendered non-significant. This compelling observation strongly indicates that, under the influence of FFM, an effective integration of weights between these two crucial components takes place. Conversely, in the absence of FFM, our discernment exposes a noteworthy pattern wherein the value of $\lambda$ exceeds 1. This conspicuous finding implies a discernible imbalance, with the weight attributed to the main body features surpassing that assigned to the boundary features. Such a circumstance underscores a heightened and more substantial contribution of the main body features to the ultimate outcome, thereby emphasizing the pivotal role of FFM in harmonizing the influence of both components in the parameter-tuning process.

\section{Discussion}

In order to rigorously evaluate our proposed model's efficacy, we conducted a comparative evaluation with various CNN-based and Transformer-based models. We demonstrate in Table 2 and Table 3 that our approach outperforms established state-of-the-art models in ultrasound image segmentation.

It is imperative to underscore that, despite the outstanding results achieved, Transformer-based methods exhibited a comparatively diminished performance in comparison to CNN-based methods. This discrepancy can be ascribed to several factors. Firstly, Transformer-based methods conventionally demand extensive datasets for training purposes, as underscored in literature \cite{43}. In the context of ultrasound image segmentation, the dataset size is relatively modest, potentially limiting the capacity of Transformer architectures to unfold their full potential. Furthermore, our employment of identical training settings for both Transformer-based and CNN-based models might not be optimal for the distinctive characteristics of the Transformer architecture, as highlighted by prior research \cite{44}\cite{45}\cite{46}.

In light of the insights gleaned from the results presented in Fig. \ref{fig7}, our model exhibits a lower Area Under the Curve (AUC) compared to the Transformer-based model. This discrepancy is attributed to the integration of numerous shallow features in our model, potentially laden with superfluous noise and extraneous information. Consequently, this may impede the model's discernment between relevant and irrelevant features, thereby influencing the AUC. Nevertheless, these findings serve as a catalyst for further enhancements in our model, advocating for the exploration of alternative feature selection techniques. These considerations offer valuable insights for future research directions in this burgeoning area.

\section{Conclusion}
This study introduces DBF-Net, a novel dual-branch network with body and boundary supervision and feature fusion for ultrasound image segmentation. Our approach incorporates a novel Feature Fusion and Supervision (FFS) module aimed at enhancing segmentation performance in ultrasound images by integrating body and boundary information and facilitating their interaction. We thoroughly evaluate DBF-Net against several state-of-the-art deep learning-based segmentation methods on challenging datasets: BUSI, UNS, and UHES. The experimental results demonstrate that DBF-Net outperforms existing approaches in ultrasound image segmentation, highlighting its potential to advance the state-of-the-art in this field.

\section*{Conflict of interest}
This work has not been published and has not been submitted for publication elsewhere while under consideration. The authors declare no potential conflict of interest.

\section*{Acknowledgments}
This work is supported by the Guangdong Provincial Key Laboratory of Human Digital Twin (No. 2022B1212010004), the Fundamental Research Funds for the Central Universities of China (No. PA2023IISL0095), and the Hubei Key Laboratory of Intelligent Robot in Wuhan Institute of Technology (No. HBIRL 202202). 

%\nolinenumbers 





\end{document}